\pdfoutput=1
\documentclass[twocolumn,showpacs,preprintnumbers,amsmath,amssymb,prl]{revtex4-1}
\usepackage{graphicx}% Include figure files
\usepackage{dcolumn}% Align table columns on decimal point
\usepackage{bm}% bold math
\begin{document}
\title{Hyperchaotic Intermittent Convection in a Magnetized Viscous Fluid}
\author{Wies{\l}aw M. Macek$^{*}$}
\affiliation{%
Faculty of Mathematics and Natural Sciences, 
Cardinal Stefan Wyszy\'{n}ski University, W\'{o}ycickiego 1/3, 01-938 Warsaw, Poland, %}%
%\altaffiliation[Also at ]{% 
and Space Research Centre, Polish Academy of Sciences, Bartycka 18A, 00-716 Warsaw, Poland}%
\email{macek@cbk.waw.pl}
\homepage{http://www.cbk.waw.pl/~macek}
\author{Marek Strumik}%
\email{maro@cbk.waw.pl}
\affiliation{%
Space Research Centre, Polish Academy of Sciences, Bartycka 18A, 00-716  Warsaw, Poland}%
\begin{abstract}
We consider a low-dimensional model of convection in a horizontally
magnetized layer of a viscous fluid heated from below. 
We analyze in detail the stability of hydromagnetic convection 
for a wide range of two control parameters. 
Namely, when changing the initially applied temperature difference or magnetic field strength,
one can see transitions from regular to irregular long-term  behavior of the system,  
switching between chaotic, periodic, and equilibrium asymptotic solutions. 
It is worth noting that owing to the induced magnetic field 
a transition to hyperchaotic dynamics 
is possible for some parameters of the model.
We also reveal new features of the generalized Lorenz model, 
including both type I and III intermittency.
\end{abstract}
\maketitle
The nature of convection in a viscous fluid  
is still not sufficiently understood. 
As well known for a fluid  heated from below in a gravitational field 
with a vertical temperature gradient,
starting from basic hydrodynamic equations 
Lorenz obtained three nonlinear ordinary differential equations \cite{Lor63}.
Following this seminal paper further studies 
have revealed the complexity of nonperiodic deterministic flow,
including strange attractors, bifurcations, 
chaotic behavior, and intermittency
(see, e.g.,  Ref. \cite{Spa82} for review). We have %recently 
generalized this model for a magnetized fluid, 
with a new variable responsible for the induced magnetic field \cite{MacStr10}.  

Transitions from regular (periodic) to irregular (nonperiodic) behavior often occur in dynamical
systems through an intermittency scenario, where signals alternate
between regular (laminar) phases and irregular bursts. 
Based on different characteristic dynamical behavior, 
three basic types I, II, and III of intermittency have been classified \cite{PomMan80},
which are related to saddle-node, Hopf, and inverse period doubling bifurcations, correspondingly.
In principle, these types of intermittent behavior can be identified experimentally
by investigating their different statistical properties. 
%More recently other intermittency mechanisms have also been found, including, e.g., 
%on-off intermittency \cite{Plaetal93}, 
%eyelet intermittency \cite{Piketal97}, 
%and ring intermittency \cite{Hraetal06}. 
In this Letter we discuss type I intermittency, 
which we have identified in the generalized Lorenz model of hydromagnetic convection, 
besides type III intermittent behavior reported
earlier in our previous paper \cite{MacStr10}.

Hyperchaos is typically defined as a complex nonperiodic 
behavior, where at least two Lyapunov exponents are positive
in contrast to standard chaotic dynamics that is characterized by
one positive Lyapunov exponent \cite{Ros79,Rosetal07}. 
Obviously, hyperchaos cannot occur in the standard Lorenz model 
because it is only possible in at least four-dimensional systems. 
In this Letter for the first time we identify such a behavior 
in our new model for hydromagnetic convection.

In general, evolution of a viscous magnetized fluid is described by the following %set of 
partial differential equations:
\begin{equation}
\frac{\mathrm{d} \mathbf{v}}{\mathrm{d} t}  =
 - \frac{1}{\rho} \mathbf{\nabla} \Big(p + \frac{\mathbf{B}^2}{2 \mu_0}\Big)
 + \frac{(\mathbf{B} \cdot \mathbf{\nabla)} \mathbf{B}}{\mu_0 \rho} 
 + \nu \triangle \mathbf{v} + \mathbf{f},
\label{e:ms:nse}
\end{equation}
\begin{equation}
\frac{\mathrm{d} \mathbf{B}}{\mathrm{d} t}  =
(\mathbf{B} \cdot \mathbf{\nabla)} \mathbf{v} + \eta \triangle \mathbf{B}, 
\label{e:ms:mfe}
\end{equation}
\begin{equation}
\frac{\mathrm{d} T}{\mathrm{d} t}  = \kappa \triangle \mathit{T},
\label{e:ms:he}
\end{equation}
where $\nu$, $\eta$, and $\kappa$ denote kinematic viscosity,  
magnetic diffusive viscosity (resistivity), and thermal conductivity of the fluid, 
in the Navier-Stokes, the magnetic advection-diffusion, 
and the heat conduction equations, respectively \cite{Lanet84}. 
This hydromagnetic problem is rather complex since 
both time and space changes, $\frac{\mathrm{d}}{\mathrm{d} t}
\equiv \frac{\partial}{\partial t} + \mathbf{v} \cdot \mathbf{\nabla}$,
of the velocity $\mathbf{v}$ of the flow, the temperature $T$ 
(with mass density $\rho$ and pressure $p$), 
and the magnetic field $\mathbf{B}$ are considered. 

We consider here a standard scenario of the Rayleigh-B\'{e}nard problem
\cite{Ray16}, a horizontal ($x$ axis) viscous fluid layer of height $h$   
heated from below with an applied vertical ($z$ axis) temperature
gradient $\delta T$ and gravitational
acceleration $g$ (see, e.g., Refs. \cite{Beret84,Sch88}). The fluid is
treated as incompressible,  $\mathbf{\nabla} \cdot \mathbf{v} = 0$,
except for the volume expansion in $f=\rho g$ term, where $\rho =
\rho_o (1 - \beta \delta T)$ (%standard Oberbeck-
Boussinesq approximation \cite{Bou1903}).

We have argued that in the case of a thin horizontal layer,
the influence of an external horizontal magnetic field $\mathbf{B_0}$ 
(along the $x$ direction) should be important \cite{MacStr10}.
Using an approximation  
${(\mathbf{B} \cdot \mathbf{\nabla)} \mathbf{v}} \approx
{(\mathbf{B_0} \cdot \mathbf{\nabla)} \mathbf{v}}$
in Eq.~(\ref{e:ms:mfe}), $\mathbf{\nabla} \cdot \mathbf{B} = 0$,
we have obtained from the general magnetohydrodynamic 
\mbox{Eqs.~(\ref{e:ms:nse})--(\ref{e:ms:he})} %a simple set of 
four ordinary differential equations \cite{MacStr10}:
\newpage
\begin{equation}
\dot{X} = -\sigma X + \sigma Y - \omega_0 W, 
\label{e:ms:x}
\end{equation}
\begin{equation}
\dot{Y} = -X Z + r X - Y, 
\label{e:ms:y}
\end{equation}
\begin{equation}
\dot{Z} =  X Y - b Z,
\label{e:ms:z}
\end{equation}    
\begin{equation}
\dot{W} = \omega_0 X - \sigma_{\mathrm m} W,
\label{e:ms:w}
\end{equation}
where a dot denotes an ordinary derivative with respect to the
normalized time \mbox{$t' = (1 + a^2) \kappa (\pi/h)^2~t$}.  
As usual $r = R_{\mathrm a} / R_{\mathrm c} $ is a
control parameter of the system  proportional to the temperature
gradient $\delta T$,  or a Rayleigh number $R_{\mathrm a} = g \beta h^3
\delta T / (\nu \kappa)$  normalized by a critical number $R_{\mathrm c}
= (1 + a^2)^3(\pi^2/a)^2$. 

In the standard three-dimensional Lorenz model,  
a time-dependent variable $X$ is proportional to the intensity of the convective motion,  
$Y$ and $Z$ describe the temperature profile  
in Eq.~(\ref{e:ms:he}), 
in a double asymmetric [parameters $a$, $b = 4/(1 + a^2)$] 
Fourier representation~\cite{Lor63}.
In the generalized Lorenz model we have in addition 
a new time dependent variable $W$ 
describing the profile of the magnetic field 
induced in the convected magnetized fluid
according to Eqs.~(\ref{e:ms:nse}) and (\ref{e:ms:mfe}),
see Ref.~\cite{MacStr10}.
We have also introduced another control parameter proportional to 
the initial magnetic field strength $B_0$ applied to the system,  
which is defined here as a basic dimensionless magnetic frequency 
$\omega_0 = {v_{\mathrm A0}}/v_0$, with 
${v_{\mathrm A0}} = {{B_0}} / {(\mu_0 \rho)^{1/2}}$
and $v_0 = 4 \pi \kappa / (a b h)$. 
The last term in Eq.~(\ref{e:ms:x}) comes from 
the anisotropic tension of the magnetic field 
${(\mathbf{B} \cdot \mathbf{\nabla)} \mathbf{B}}/({\mu_0 \rho})$
in Eq.~(\ref{e:ms:nse}).
Naturally, besides the Prandtl number $\sigma = \nu/\kappa$, 
the properties of the magnetized fluid are characterized by an analogue
parameter $\sigma_{\mathrm m} = \eta/\kappa = \sigma/Pr_\mathrm{m}$ (where
$Pr_\mathrm{m}=\nu/\eta$ is the magnetic Prandtl number)
appearing now in Eq. (\ref{e:ms:w}),
and resulting from the last terms in Eqs.~(\ref{e:ms:mfe}) and (\ref{e:ms:he}).

\begin{figure}[!htbp]
\includegraphics[height=65mm]{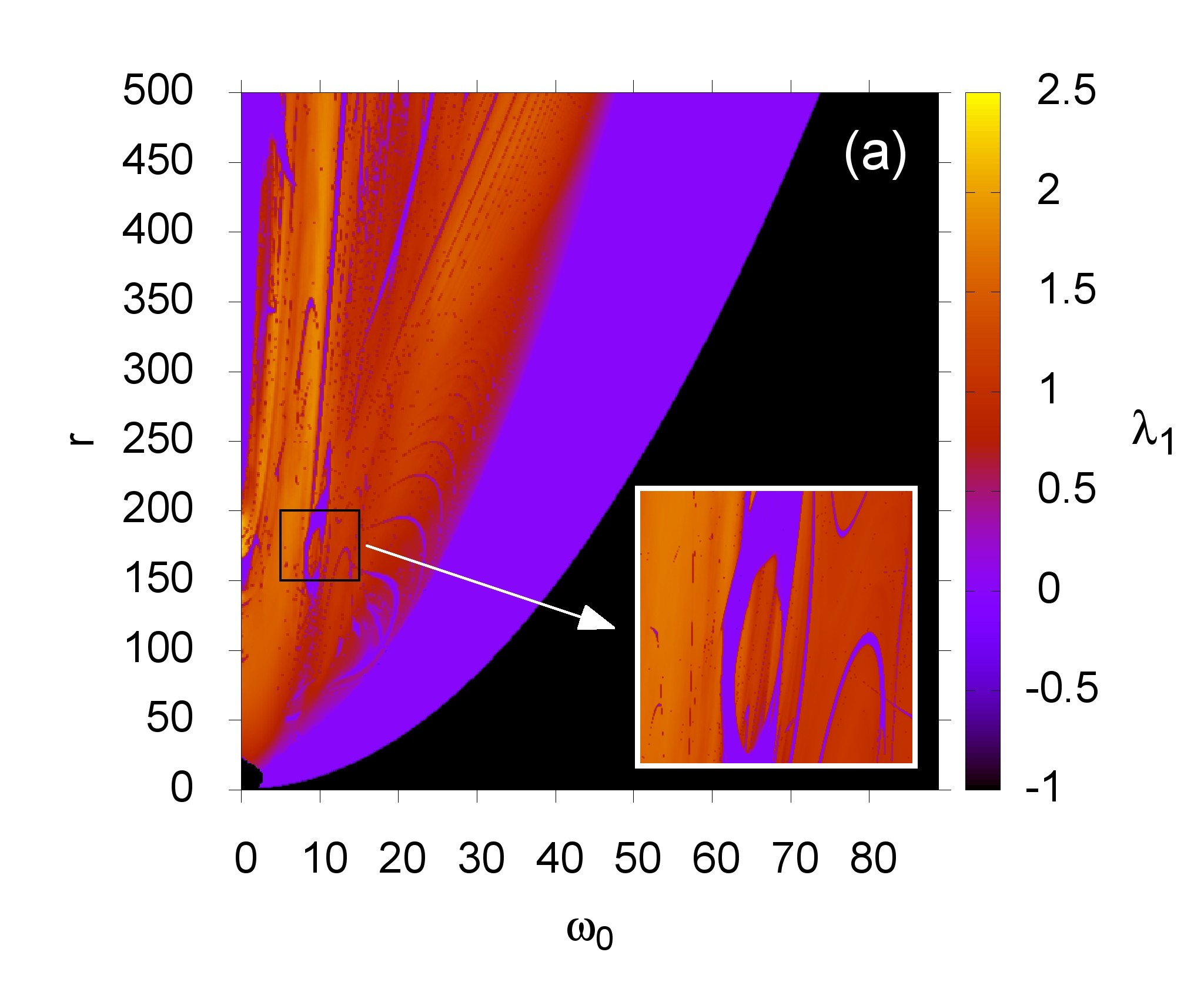}
\includegraphics[height=65mm]{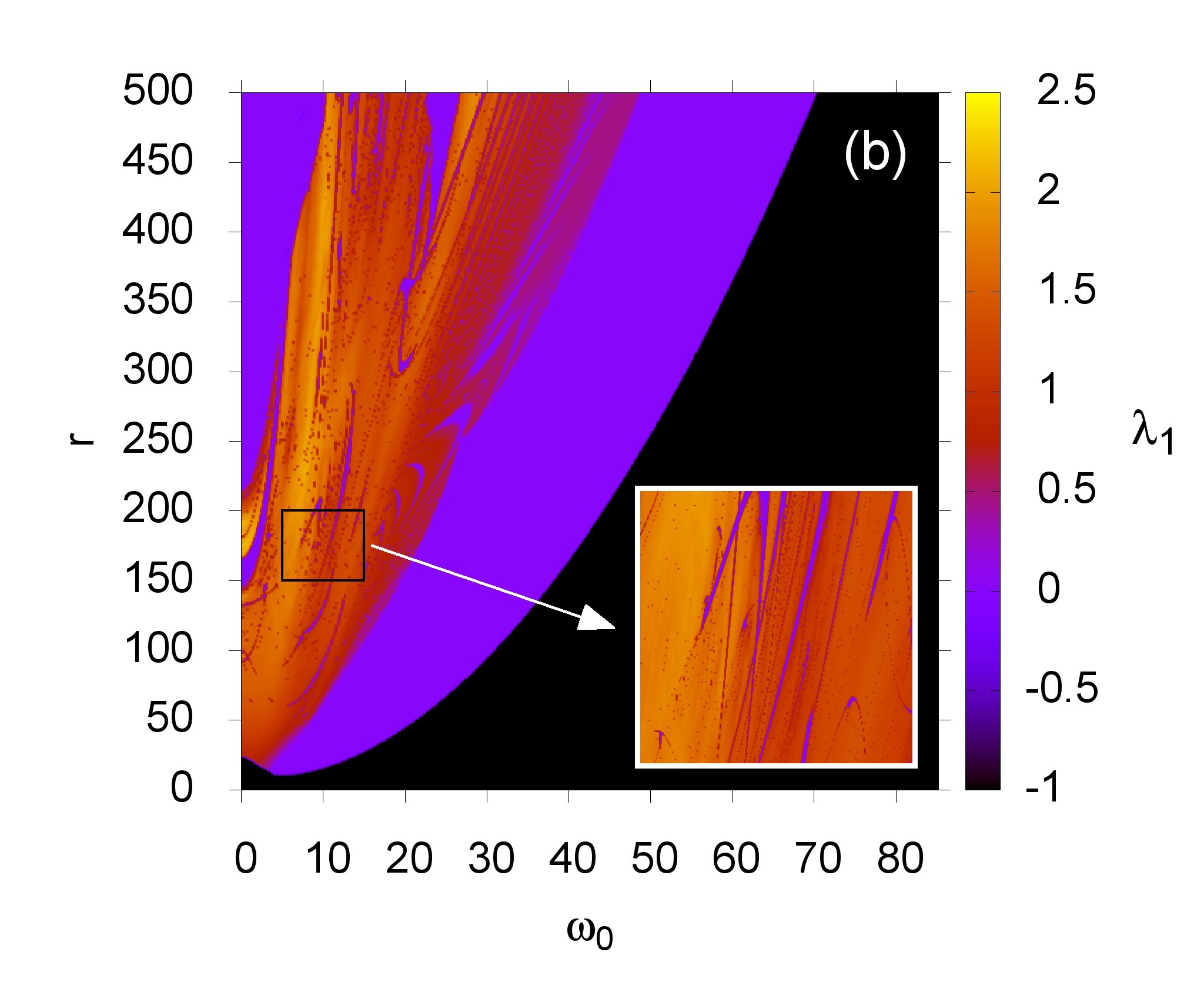}
\includegraphics[height=65mm]{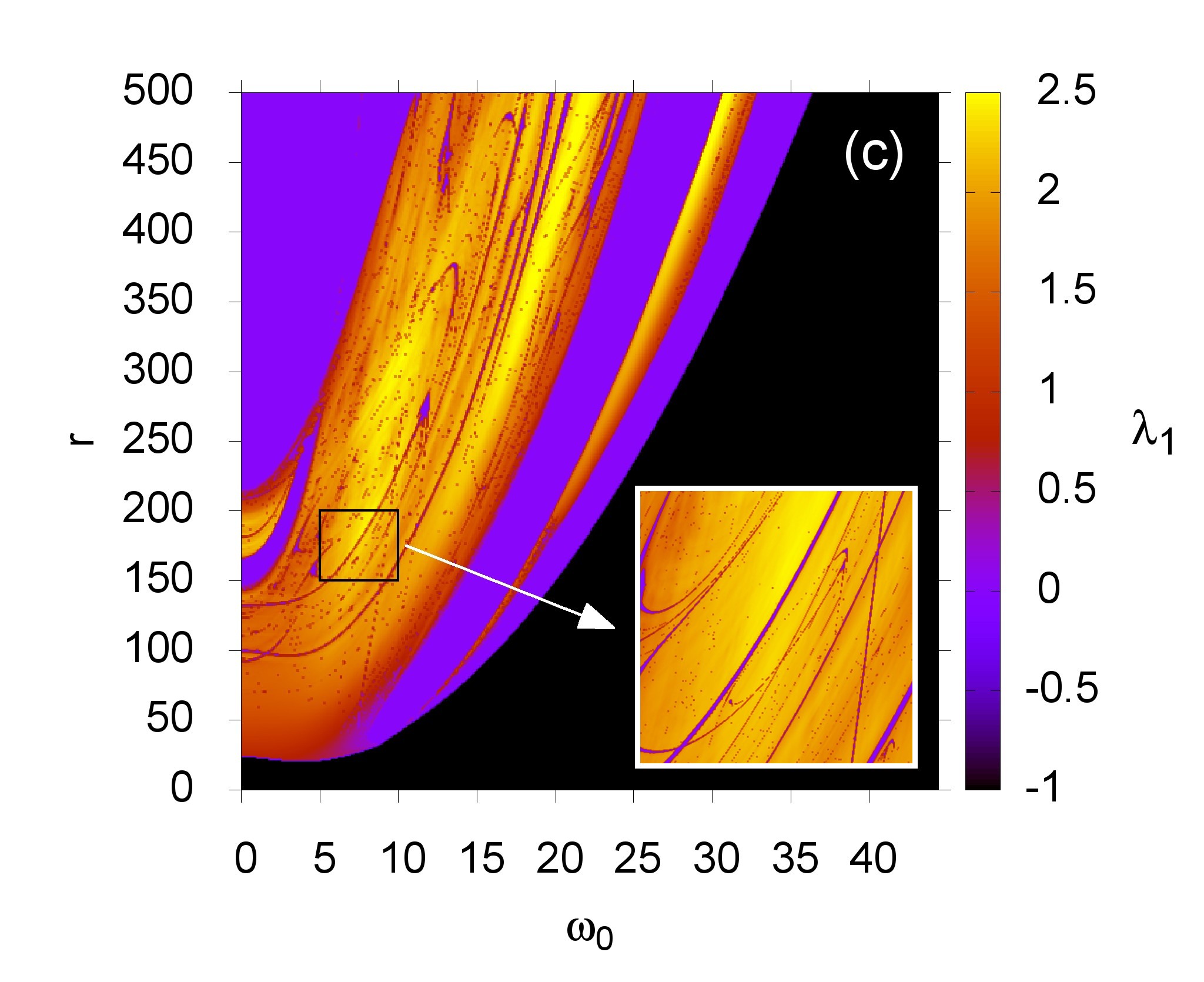}
\caption{
Dependence of the largest Lyapunov exponent $\lambda_1$ (color coded) on
$\omega_0$ and $r$ parameters of the generalized Lorenz model for
(a) $\sigma_{\mathrm m} = 0.1$, (b) $\sigma_{\mathrm m}= 1$, and
(c) $\sigma_{\mathrm m} = 3$.  Other parameters of the
system have fixed values: $\sigma = 10$, $b = 8/3$. 
Convergence of the solutions 
of Eqs.~(\ref{e:ms:x})--(\ref{e:ms:w}) to fixed points ($\lambda_1<0$) is
shown in black, to periodic solutions ($\lambda_1=0$) --- in violet/blue color
(see the color bar for $\lambda_1=0$), to chaotic solutions ($\lambda_1>0$) ---
in a color, consistently with the color bar scale, from violet/blue to yellow.
%For every panel (a)-(c) an enlargement of the region 
%bounded by black lines is shown in the right-bottom part of plots.
Fine structures are shown in the insets.
}
\label{f:lyap_anal}
\end{figure}

In Fig. \ref{f:lyap_anal} we present plots of the largest Lyapunov exponent
illustrating long-term (asymptotic)
behavior of the dynamical system of \mbox{Eqs.~(\ref{e:ms:x})--(\ref{e:ms:w})}
in the space of dimensionless control parameters $\omega_0$ and $r$.
The Lyapunov exponents are computed for solutions of \mbox{Eqs.~(\ref{e:ms:x})--(\ref{e:ms:w})}
using the QR decomposition method (discussed thoroughly in Sec.~VC of Ref.~\cite{EckRue85}) 
that provides reliable and accurate estimation 
of the full spectrum of the exponents 
when differential equations are explicitly known.
The method requires long time series 
(that naturally appear in the case of periodic or chaotic solutions); 
thus a preparatory step has been applied 
to detect convergence of a given solution to a fixed point. 
%Depending on the sign (and values) of the calculated largest Lyapunov exponent,
%various regions of convergence are shown in different colors.
%Namely, fixed points (lack of long-term limit cycle oscillations, negative exponent) 
%are indicated -- in black, periodic solutions (zero value) -- in violet, 
%and chaotic dynamics (with positive exponent) -- in a color,
%consistently with the color bar scale, from violet to yellow,
%correspondingly. 
Three cases of parameter  $\sigma_{\mathrm m}$ (ratio of  Prandtl numbers) 
are considered here as related to different magnitude of resistive dissipation affecting the system.

In Ref. \cite{MacStr10} a single case was studied parametrically:
$\sigma_{\mathrm m} = 1$ and $0 < r < 50$, 
which corresponds to the bottom left part of Fig.~\ref{f:lyap_anal}(b).
A range of $0 < r <500$ in the present study is ten times larger 
as compared with Ref.~\cite{MacStr10}, 
which allows us to identify new features of the generalized Lorenz model. 
The present extended analysis shows that plots for different values of
$\sigma_{\mathrm m}$ have roughly similar structure. 
In the right bottom part (where long-term limit cycle oscillations are not possible)
solutions converge to fixed points (corresponding to equilibria). 
Next in the proximity of the diagonal a wide region of periodic (limit cycle) behavior is located, 
and further toward the left top part of the plots one may find nonperiodic (chaotic) solutions 
followed by periodic region close to the left top corner. 
For $\sigma_{\mathrm m} = 0.1$ and $1$ [Figs.~\ref{f:lyap_anal}(a) and (b)] 
the ``diagonal'' region of periodic solutions is homogeneous (without gaps) 
whereas for $\sigma_{\mathrm m}=3$ [Fig.~\ref{f:lyap_anal}(c)] 
one can see a gap of chaotic solutions dividing this region into two parts. 
More detailed inspection of the plots reveals complicated structure of regions, 
where domains of chaotic solutions are intertwined finely with domains of periodic solutions 
that makes influence of the parameters $\omega_0$ and $r$ even more complicated,
%as seen in the enlargements shown in the right-bottom part of the plots in 
see insets to Fig.~\ref{f:lyap_anal}.
The fine structure in the parameter space is seen for all values of 
$\sigma_{\mathrm m}$ parameter considered here, 
and it implies interesting properties of the dynamics 
as regards to regularity of convective motions,
when affected by changing boundary conditions related to 
control parameters $r$ and $\omega_0$ used in the model.

%Naturally, the control parameter $\omega_0$ 
%is related to initial magnetic field $B_0$ applied to the dynamical system, 
%thus dependence of solutions on $\omega_0$ allows us to study 
%the influence of external constant magnetic field on the dynamics. 
As seen in Fig.~\ref{f:lyap_anal} the dependence of solutions
of the system on $\omega_0$ parameter (related to $B_0$) can be quite complicated. 
In the simplest case, e.g., for $r = 50$, 
ignoring the fine structure of the chaotic region, 
for the increasing $\omega_0$ parameter 
we observe a transition from chaotic (through periodic) to fixed point solutions, 
cf. (Fig.~1 in \cite{MacStr10}). 
This suggests purely damping influence 
of the increasing external magnetic field. 
However, %as seen here 
for higher values of $r$, e.g., for $r =  350$, 
we rather observe more complex transitions between dynamical regimes: 
``periodic'' -- ``chaotic'' -- ``periodic'' -- ``fixed point'' 
for $\sigma_\mathrm{m} \leq 1$ [Figs.~\ref{f:lyap_anal}(a) and (b)] 
and ``periodic'' -- ``chaotic'' -- ``periodic'' -- ``chaotic'' -- ``periodic'' -- ``fixed point'' 
for $\sigma_\mathrm{m} = 3$ [Fig.~\ref{f:lyap_anal}(c)]. 

The external temperature gradient $\delta T$ also
influences the dynamics in an intricate manner 
as one can see in Fig.~\ref{f:lyap_anal} analyzing the dependence of solutions 
on $r \propto \delta T$ for fixed values of $\omega_0$. 
Obviously, for $\omega_0 = 0$ we obtain a description fully corresponding 
to that for classical Lorenz equations of Ref.~\cite{Lor63}
with the well-known Lorenz strange attractor.
This can be easily understood by analysis of
Eqs.~(\ref{e:ms:x})--(\ref{e:ms:w}), where Eq.~(\ref{e:ms:w}) 
decouples from the first three equations for $\omega_0 = 0$, 
which gives classical Lorenz system. 
In this case the dynamical scenario for increasing $r$ with $\omega_0 = 0$ 
can be described as the following transition between dynamical regimes: 
``fixed point'' -- ``chaotic'' -- ``periodic'' -- ``chaotic'' -- ``periodic''. 
However, for higher values of $\omega_0$, e.g., $\omega_0 = 15$ 
the initial transition from fixed point to chaotic dynamics 
has an intermediate phase of limit cycle periodic oscillations.
Some new strange attractors appearing for %in the case of 
the magnetized fluid have been presented in Ref. \cite{MacStr10}. 

\begin{figure}[!t]
\includegraphics[height=70mm]{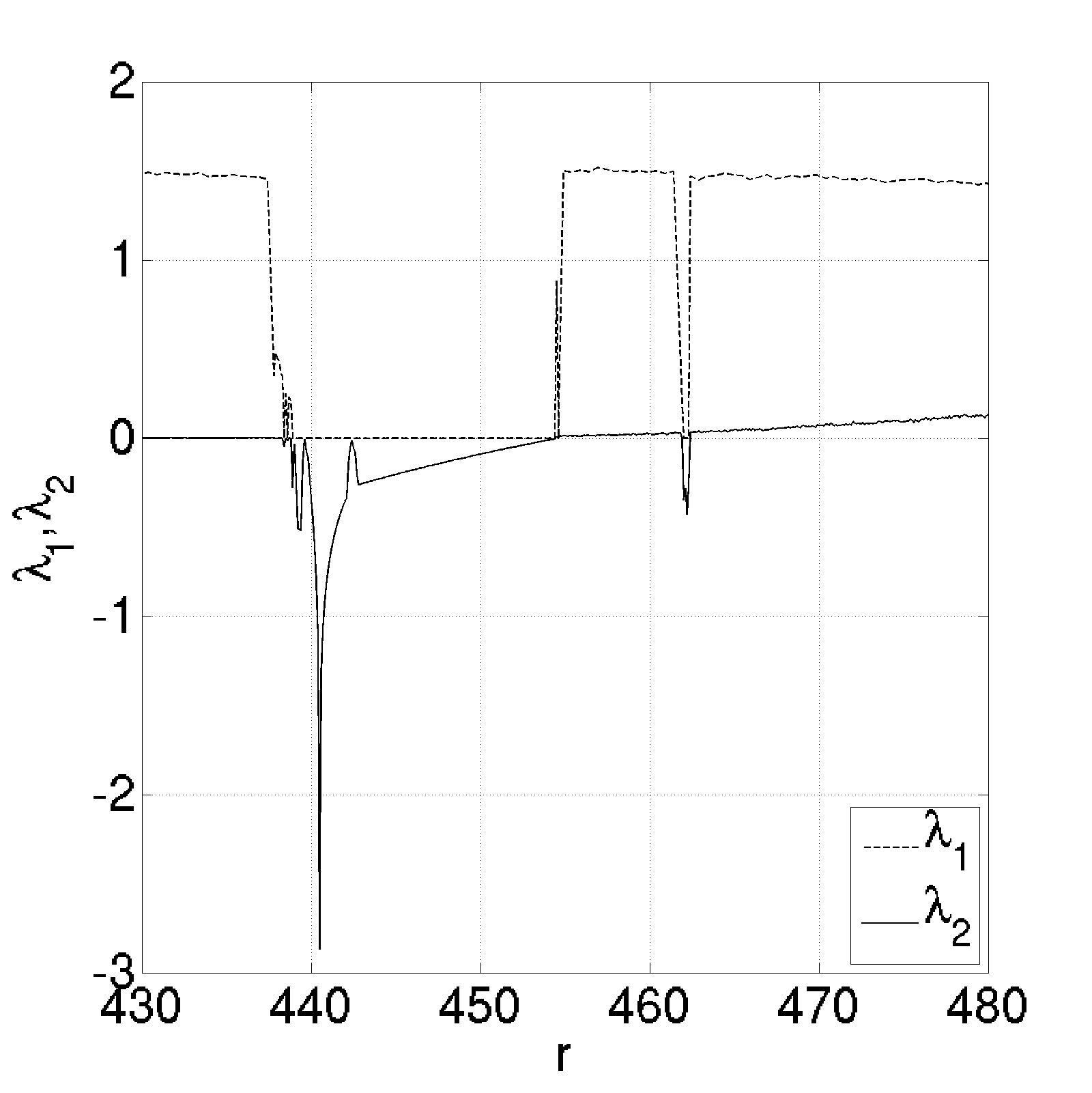}
\caption{Dependence of the 
two largest Lyapunov exponents
$\lambda_1$ (dashed line) and $\lambda_2$ (solid line), $\lambda_1>\lambda_2$, on parameter $r$. 
%of the system for fixed values of the other parameters.
%: $\omega_0=5.95$, $\sigma_{\mathrm m}=0.1$, $\sigma=10$, and $b=8/3$. 
%A transition to hyperchaotic dynamics is observed, 
%when the second Lyapunov exponent $\lambda_2$ becomes positive for $r \ge 454.7$.
}
\label{f:hyper}
\end{figure}

It is even more interesting that in the range \mbox{$0< r < 500$} %considered in this paper 
we have identified hyperchaotic solutions for small values of $\sigma_\mathrm{m}$ parameter. 
Namely, in Fig.~\ref{f:hyper}
we show the dependence of the two largest Lyapunov exponents 
$\lambda_1$ (dashed line) and $\lambda_2$ (solid line), $\lambda_1>\lambda_2$,
on the \mbox{parameter} $r$ of the system for $\omega_0 = 5.95$, $\sigma_{\mathrm m} = 0.1$
with fixed values of the other standard Lorenz system parameters $\sigma = 10$, and $b = 8/3$. 
The second Lyapunov exponent $\lambda_2$ becomes positive for $r \ge 454.7$, which %exhibits 
implies a transition to hyperchaotic dynamics. 
The largest Lyapunov exponent $\lambda_1$ increases abruptly during this transition,
whereas the second exponent $\lambda_2$ also becomes positive still increasing its value rather smoothly. 
The region of hyperchaotic behavior has a gap for $461.8< r <462.4$,
where periodic solutions appear, %as seen in 
see Fig.~\ref{f:hyper}.
These results may be of special interest for experimental
identification of hyperchaotic dynamics in plasmas. 
Admittedly, it could rather be difficult to identify such a system in general  
because any dynamical system exhibiting divergence of trajectories in two directions 
(with two positive Lyapunov exponents) 
is clearly more complex than a chaotic system with only one such an unstable direction
%Additionally, analysis of real systems is typically based on measurements of scalar time series, 
%which may cause significant observational problems, 
%when this time series is used to infer about four (and more) dimensional dynamics. 
%These observability problems usually result from spurious folding 
%appearing in projection from four (and more) dimensional space 
%to one-dimensional space and are difficult to cope with by using the embedding procedure, 
%as discussed thoroughly, e.g. in Ref. 
\cite{Rosetal07}. 
In this context, analysis of statistical properties 
(e.g. distributions or scaling in intermittency) of the observed dynamical behavior  
can be more interesting from experimental point of view; 
thus we discuss the statistics below. 

In Ref.~\cite{MacStr10} some solutions (for $r = 28$, $\omega_0=4.8$,
$\sigma_\mathrm{m} = 1$) of the dynamical system of Eqs.~(\ref{e:ms:x})--(\ref{e:ms:w}) 
have been discussed as examples of type III intermittent behavior. 
In the present study we extend this analysis to other cases. It is %well 
known that the classical Lorenz system exhibits type I intermittency
transition from periodic to chaotic dynamics for the value of control parameter $r \approx 166.06$. 
In fact, in Fig.~\ref{f:lyap_anal} one can see a branch of periodic-chaotic boundary 
originating from this point for $\omega_0 = 0$ in the parameter plane. 
When the magnetic field is taken into account,
type I intermittency occurs along this branch, e.g., 
for $r=256$, $\omega_0\approx3.74$, $\sigma_\mathrm{m}=1$. 

\begin{figure}[!tbp]
\includegraphics[height=50mm]{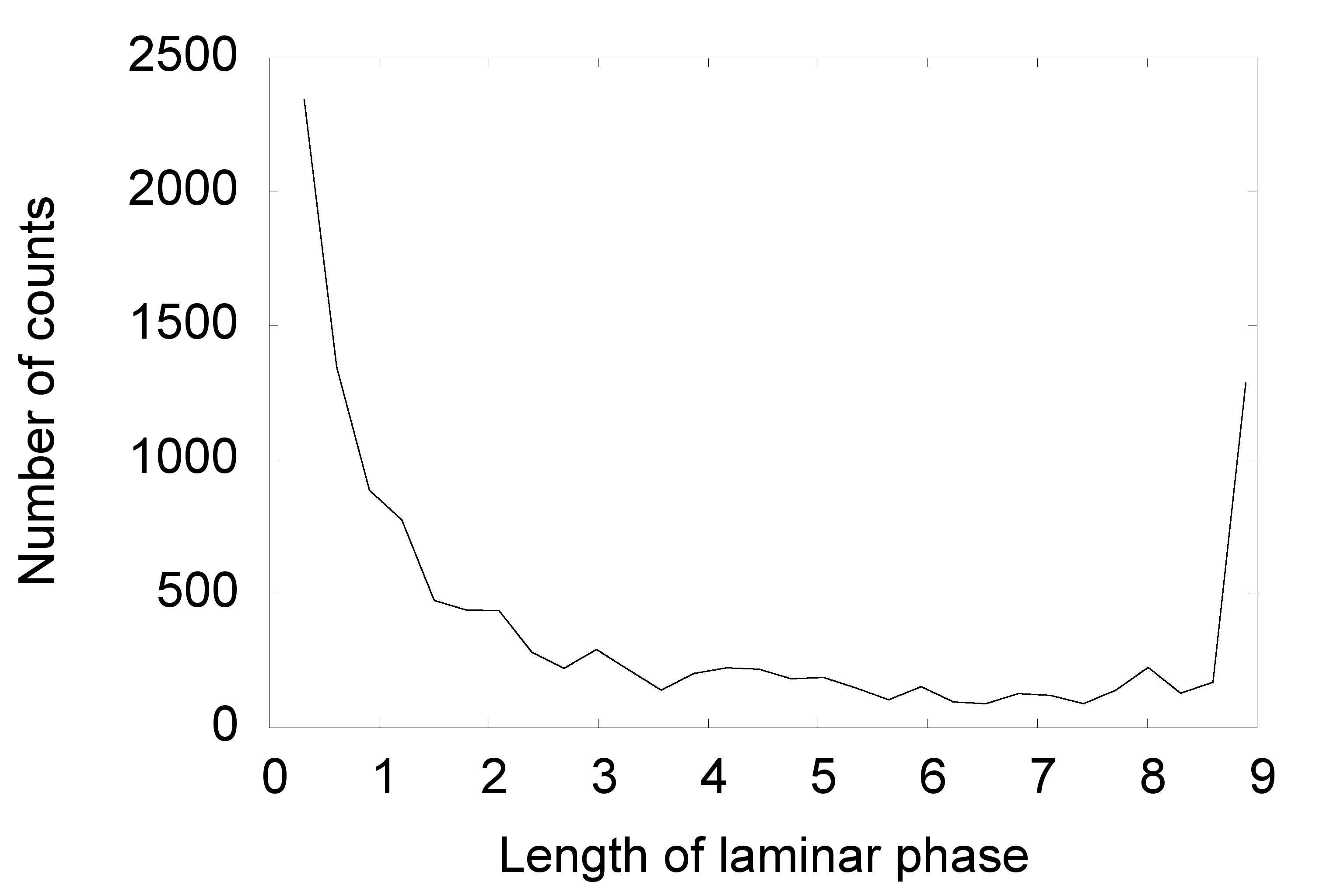}
\caption{Distribution of the lengths of laminar phases for
$r=256$, $\omega_0= 3.74$, $\sigma_\mathrm{m}=1$. 
%The distribution is characterized by U-shape and finite value of maximum length 
%of laminar phase, which is characteristic for type I intermittency.
}
\label{f:intermit1distr}
\end{figure}

\begin{figure}[!tbp]
\includegraphics[height=50mm]{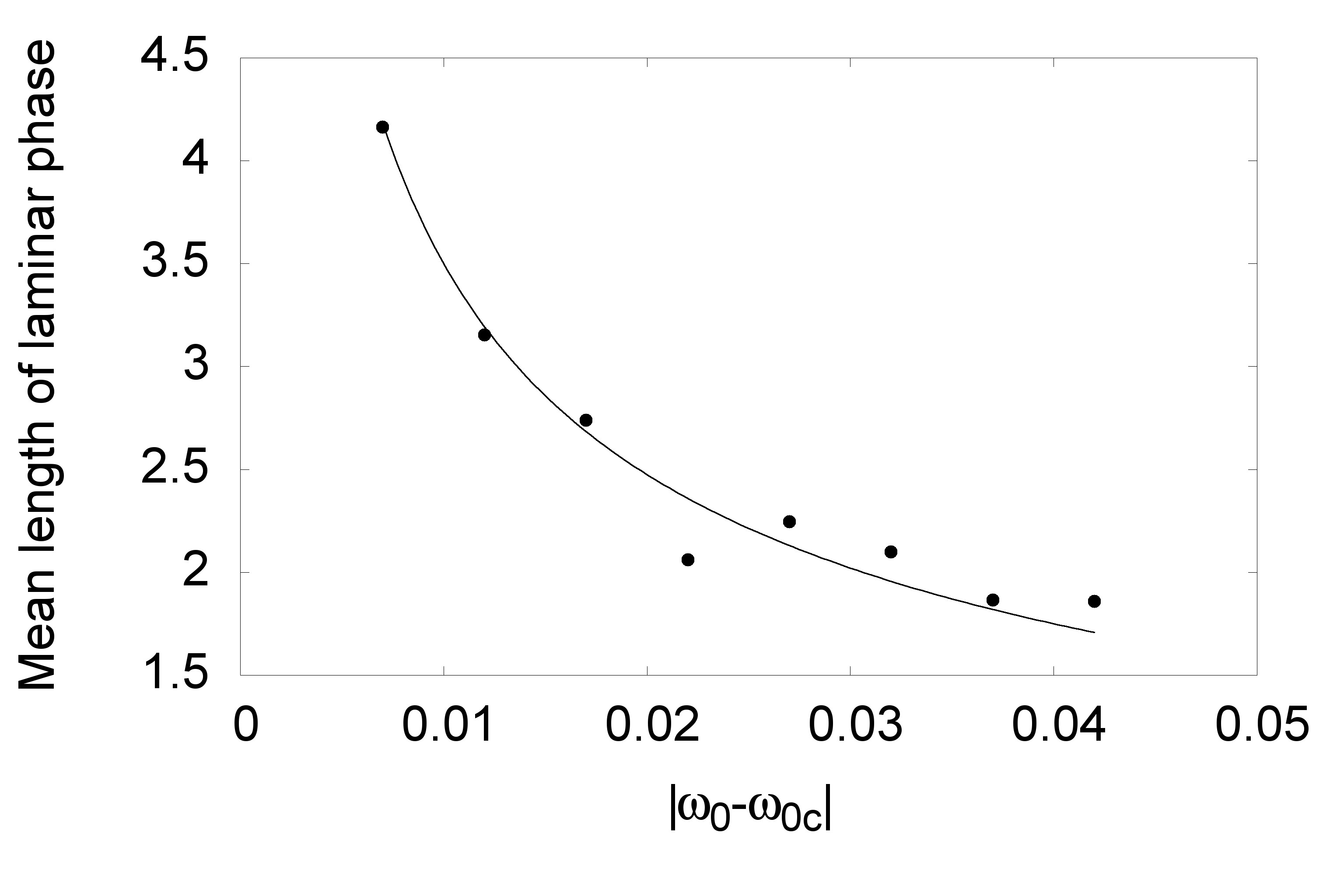}
\caption{Scaling of the mean length of the laminar phase 
with control parameter $\epsilon = |\omega_0 - \omega_{0c}|$ for
the case shown in Fig.~\ref{f:intermit1distr},
where $\omega_{0c}$ is a critical value at which intermittency appears. 
%The dependence resulting from computations (circles) can be  approximated by
%$\propto \epsilon ^{-1/2}$ function (solid line) 
%characteristic for type I intermittency.
}
\label{f:intermit1scal}
\end{figure}

Next, we determine the lengths of laminar phases and their distribution  
using an algorithm, where pieces of a long numerical solution are compared to 
a periodic (laminar) phase pattern in four-dimensional phase space.
The piecewise numerical solution of Eqs.~(\ref{e:ms:x})--(\ref{e:ms:w})
is a set of points in the phase space, thus based on the average
distance between the points and their nearest neighbors found in the laminar pattern 
we can identify laminar phases. 
As demonstrated in Fig.~\ref{f:intermit1distr}
for the type I intermittency 
with characteristic $U$ shape of the distribution of laminar phases
the maximum length of laminar phase has some finite value. 
Moreover, as shown in Fig.~\ref{f:intermit1scal} 
in this case we observe another characteristic attributes of the type I intermittency, 
namely scaling of the mean length of laminar phase with control parameter 
$\propto \epsilon^{-1/2}$, where $\epsilon = |\omega_0 - \omega_{0c}|$.

In conclusion, the four-dimensional dynamical system for convection in a magnetized viscous fluid
exhibits quite unusual features depending on the control parameters of the model.
It is known that increasing temperature gradient $\delta T$ 
in the standard Lorenz equations does not imply more chaos in the model \cite{Spa82}. 
Our study provides detailed picture of this counterintuitive influence of 
$\delta T$ on the generalized Lorenz model, 
where the system goes through intertwined regions
of chaotic, periodic, and fixed-point asymptotic solutions. 
Quite surprisingly a similar complicated influence is seen for
systematic increase of the background magnetic field strength $B_0$. 
The fine structure %illustrated 
in the control parameters space clearly shows
that the influence of $B_0$ %this background magnetic field 
is much more intricate than a simple stabilizing effect 
predicted by simplified analysis of influence of the magnetic field 
on convective motion discussed in textbooks (see, e.g. Ref.~\cite{Cow76}). 
This is interesting because physical circumstances are indeed known 
where even weak field may have strong destabilizing effect \cite{BajMiz13}. 

In a chaotic regime but near the border with periodic solutions,
in addition to previously identified type III intermittency,
we have also observed type I intermittent behavior of the system
that could provide new mechanisms of release of kinetic and magnetic energy bursts.
It is worth noting that the observed sudden transitions from regular to irregular behavior 
only mimic stochastic forces, but in fact they result from nonlinearity; i.e.,
they are due to the disappearance of the fixed points of the dynamical system 
or owing to a change in their stability.

It is important to note that besides the chaotic behavior well known for the Lorenz model with unmagnetized fluid,
we have also identified here for the first time a hyperchaotic dynamics in a magnetic dynamical system, 
with two positive Lyapunov exponents appearing for some value of the intensity of the applied magnetic field. 
Admittedly, this new type of chaos is only possible  in at least four-dimensional system,
hence this results here from the interplay between anisotropic tension of magnetic field lines and  magnetic viscosity. 

Basically, our analysis focuses on characteristic signatures of the hydromagnetic convection 
that can be relevant for observational identification of this kind of dynamical behavior, 
e.g., through analysis of statistical properties of observed intermittent energy bursts 
as compared with those predicted by the model presented in this Letter. 
In this context the new hyperchaotic system characterized by both types I and III of intermittent energy release 
may provide an approximate description of irregular convective dynamical processes 
observed often in various magnetized plasmas in both laboratory and space,
e.g., for solar sunspots \cite{Noretal09}, planetary and stellar liquid interiors \cite{Bus00}, 
and possibly for magnetoconfined plasmas in tokamaks \cite{Antetal03}, 
nanodevices and microchannels in nanotechnology.

\newpage
This work was supported by the %Polish 
National Science \mbox{Center} (NCN) through Grant NN 307 0564 40.
W.~M. acknow\-ledges support by the European Community's Seventh Framework Programme ([FP7/2007-2013])
under Grant agreement No. 313038/STORM.

%\bibliography{ms13}

\begin{thebibliography}{17}
\expandafter\ifx\csname natexlab\endcsname\relax\def\natexlab#1{#1}\fi
\expandafter\ifx\csname bibnamefont\endcsname\relax
  \def\bibnamefont#1{#1}\fi
\expandafter\ifx\csname bibfnamefont\endcsname\relax
  \def\bibfnamefont#1{#1}\fi
\expandafter\ifx\csname citenamefont\endcsname\relax
  \def\citenamefont#1{#1}\fi
\expandafter\ifx\csname url\endcsname\relax
  \def\url#1{\texttt{#1}}\fi
\expandafter\ifx\csname urlprefix\endcsname\relax\def\urlprefix{URL }\fi
\providecommand{\bibinfo}[2]{#2}
\providecommand{\eprint}[2][]{\url{#2}}

\bibitem[{\citenamefont{{Lorenz}}(1963)}]{Lor63}
\bibinfo{author}{\bibfnamefont{E.~N.} \bibnamefont{{Lorenz}}},
  \bibinfo{journal}{J. Atmos. Sci.} \textbf{\bibinfo{volume}{20}},
  \bibinfo{pages}{130} (\bibinfo{year}{1963}).

\bibitem[{\citenamefont{{Sparrow}}(1982)}]{Spa82}
\bibinfo{author}{\bibfnamefont{C.}~\bibnamefont{{Sparrow}}},
  \emph{\bibinfo{title}{The Lorenz Equations: Bifurcations, Chaos and Strange
  Attractors}} (\bibinfo{publisher}{Springer-Verlag, Berlin},
  \bibinfo{year}{1982}).

\bibitem[{\citenamefont{{Macek} and {Strumik}}(2010)}]{MacStr10}
\bibinfo{author}{\bibfnamefont{W.~M.} \bibnamefont{{Macek}}} \bibnamefont{and}
  \bibinfo{author}{\bibfnamefont{M.}~\bibnamefont{{Strumik}}},
  \bibinfo{journal}{\pre} \textbf{\bibinfo{volume}{82}},
  \bibinfo{pages}{027301} (\bibinfo{year}{2010}).

\bibitem[{\citenamefont{{Pomeau} and {Manneville}}(1980)}]{PomMan80}
\bibinfo{author}{\bibfnamefont{Y.}~\bibnamefont{{Pomeau}}} \bibnamefont{and}
  \bibinfo{author}{\bibfnamefont{P.}~\bibnamefont{{Manneville}}},
  \bibinfo{journal}{Commun. Math. Phys.} \textbf{\bibinfo{volume}{74}},
  \bibinfo{pages}{189} (\bibinfo{year}{1980}).


\bibitem[{\citenamefont{{Rossler}}(1979)}]{Ros79}
\bibinfo{author}{\bibfnamefont{O.~E.} \bibnamefont{{Rossler}}},
  \bibinfo{journal}{Phys. Lett. A} \textbf{\bibinfo{volume}{71}},
  \bibinfo{pages}{155} (\bibinfo{year}{1979}).

\bibitem[{\citenamefont{Rossler}(2007)}]{Rosetal07}
\bibinfo{author}{  \bibfnamefont{C.~Letellier} \bibnamefont{and}
  \bibfnamefont{O.~E.} \bibnamefont{Rossler}}, 
  \bibinfo{journal}{Scholarpedia}
  \textbf{\bibinfo{volume}{2}}, \bibinfo{pages}{1936} (\bibinfo{year}{2007}).

\bibitem[{\citenamefont{{Landau} et~al.}(1984)\citenamefont{{Landau},
  {Lifshitz}, and {Pitaevskii}}}]{Lanet84}
\bibinfo{author}{\bibfnamefont{L.~D.} \bibnamefont{{Landau}}},
  \bibinfo{author}{\bibfnamefont{E.~M.} \bibnamefont{{Lifshitz}}},
  \bibnamefont{and} \bibinfo{author}{\bibfnamefont{L.~P.}
  \bibnamefont{{Pitaevskii}}}, \emph{\bibinfo{title}{Electro\-dynamics of
  Continuous Media}}, (\bibinfo{publisher}{Pergamon Press, Oxford}, 
  \bibinfo{year}{1984}), vol.~\bibinfo{volume}{8}.

\bibitem[{\citenamefont{{{Lord} {Rayleigh}}}(1916)}]{Ray16}
\bibinfo{author}{\bibnamefont{{{Lord} {Rayleigh}}}}, \bibinfo{journal}{Philos.
  Mag.} \textbf{\bibinfo{volume}{32}}, \bibinfo{pages}{529}
  (\bibinfo{year}{1916}).

\bibitem[{\citenamefont{{Berg{\'e}} et~al.}(1984)\citenamefont{{Berg{\'e}},
  {Pomeau}, and {Vidal}}}]{Beret84}
\bibinfo{author}{\bibfnamefont{P.}~\bibnamefont{{Berg{\'e}}}},
  \bibinfo{author}{\bibfnamefont{Y.}~\bibnamefont{{Pomeau}}}, \bibnamefont{and}
  \bibinfo{author}{\bibfnamefont{C.}~\bibnamefont{{Vidal}}},
  \emph{\bibinfo{title}{{Order within Chaos. Towards a Deterministic Approach
  to Turbulence}}} (\bibinfo{publisher}{Wiley, New York},
  \bibinfo{year}{1984}).

\bibitem[{\citenamefont{{Schuster}}(1988)}]{Sch88}
\bibinfo{author}{\bibfnamefont{H.~G.} \bibnamefont{{Schuster}}},
  \emph{\bibinfo{title}{{Deterministic Chaos. An Introduction}}}
  (\bibinfo{publisher}{VCH Verlagsgesellschaft, Weinheim},
  \bibinfo{year}{1988}).

\bibitem[{\citenamefont{{Boussinesq}}(1903)}]{Bou1903}
\bibinfo{author}{\bibfnamefont{J.}~\bibnamefont{{Boussinesq}}},
  \emph{\bibinfo{title}{Th{\'e}orie Analytique de la Chaleur}}
  (\bibinfo{publisher}{Gauthier-Villars}, \bibinfo{year}{1903}).

\bibitem[{\citenamefont{{Eckmann} and {Ruelle}}(1985)}]{EckRue85}
\bibinfo{author}{\bibfnamefont{J.-P.} \bibnamefont{{Eckmann}}}
  \bibnamefont{and} \bibinfo{author}{\bibfnamefont{D.}~\bibnamefont{{Ruelle}}},
  \bibinfo{journal}{Rev. Mod. Phys.} \textbf{\bibinfo{volume}{57}},
  \bibinfo{pages}{617} (\bibinfo{year}{1985}).

\bibitem[{\citenamefont{{Cowling}}(1976)}]{Cow76}
\bibinfo{author}{\bibfnamefont{T.~G.} \bibnamefont{{Cowling}}},
  \emph{\bibinfo{title}{Magnetohydrodynamics}} 
  (\bibinfo{publisher}{Adam Hilger, Bristol}, \bibinfo{year}{1976}).

\bibitem[{\citenamefont{{Bajer} and {Mizerski}}(2013)}]{BajMiz13}
\bibinfo{author}{\bibfnamefont{K.}~\bibnamefont{{Bajer}}} \bibnamefont{and}
  \bibinfo{author}{\bibfnamefont{K.}~\bibnamefont{{Mizerski}}},
  \bibinfo{journal}{Phys. Rev. Lett.} \textbf{\bibinfo{volume}{110}},
  \bibinfo{eid}{104503} (\bibinfo{year}{2013}).

\bibitem[{\citenamefont{{Nordlund} et~al.}(2009)\citenamefont{{Nordlund},
  {Stein}, and {Asplund}}}]{Noretal09}
\bibinfo{author}{\bibfnamefont{{\AA}.}~\bibnamefont{{Nordlund}}},
  \bibinfo{author}{\bibfnamefont{R.~F.} \bibnamefont{{Stein}}},
  \bibnamefont{and}
  \bibinfo{author}{\bibfnamefont{M.}~\bibnamefont{{Asplund}}},
  \bibinfo{journal}{Living Rev. Solar Phys.}
  \textbf{\bibinfo{volume}{6}}, \bibinfo{pages}{2} (\bibinfo{year}{2009}).

\bibitem[{\citenamefont{{Busse}}(2000)}]{Bus00}
\bibinfo{author}{\bibfnamefont{F.~H.} \bibnamefont{{Busse}}},
  \bibinfo{journal}{Annu. Rev. Fluid Mech.}
  \textbf{\bibinfo{volume}{32}}, \bibinfo{pages}{383} (\bibinfo{year}{2000}).

\bibitem[{\citenamefont{{Antar} et~al.}(2003)\citenamefont{{Antar}, {Counsell},
  {Yu}, {Labombard}, and {Devynck}}}]{Antetal03}
\bibinfo{author}{\bibfnamefont{G.~Y.} \bibnamefont{{Antar}}},
  \bibinfo{author}{\bibfnamefont{G.}~\bibnamefont{{Counsell}}},
  \bibinfo{author}{\bibfnamefont{Y.}~\bibnamefont{{Yu}}},
  \bibinfo{author}{\bibfnamefont{B.}~\bibnamefont{{Labombard}}},
  \bibnamefont{and}
  \bibinfo{author}{\bibfnamefont{P.}~\bibnamefont{{Devynck}}},
  \bibinfo{journal}{Phys. Plasmas} \textbf{\bibinfo{volume}{10}},
  \bibinfo{pages}{419} (\bibinfo{year}{2003}).

\end{thebibliography}

\end{document}